\documentclass[aps,prc,amsfonts,showpacs,floats,floatfix,superscriptaddress,twocolumn,nofootinbib,byrevtex]{revtex4}
\usepackage{epsfig}
\usepackage{amsmath}
\usepackage{amssymb} 
\begin{document}
\title{From Finite Nuclei to the Nuclear Liquid Drop: Leptodermous 
Expansion Based on the Self-consistent Mean-Field Theory}
\author{P.-G. Reinhard}
\affiliation{Institut f\"ur Theoretische Physik II,
             Universit\"at Erlangen--N\"urnberg,
             Staudtstrasse 7, D--91058 Erlangen, Germany} 
\affiliation{Joint Institute for Heavy Ion Research,
             Oak Ridge National Laboratory,
             P. O. Box 2008, Oak Ridge, Tennessee 37831, 
             U.S.A.}
\author{M. Bender} 
\affiliation{Institute for Nuclear Theory,
             University of Washington,
             Box 351550, Seattle, WA 98195-1550, 
             U.S.A.}
\affiliation{Physics Division,
             Argonne National Laboratory,
             9700 S. Cass Avenue,
             Argonne, IL 60439,
             U.S.A.}
\affiliation{National Superconducting Cyclotron Laboratory,
             Michigan State University, 
             East Lansing, MI 48824,
             U.S.A.}
\author{W. Nazarewicz}
\affiliation{Department of Physics and Astronomy,
             The University of Tennessee,
             Knoxville, Tennessee 37996, 
             U.S.A.}
\affiliation{Physics Division,
             Oak Ridge National Laboratory,
             P. O. Box 2008, Oak Ridge, Tennessee 37831, 
             U.S.A.}
\affiliation{Institute of Theoretical Physics,
             Warsaw University,
             ul. Ho\.za 69, PL-00681, Warsaw, Poland}
\author{T. Vertse}
\affiliation{Joint Institute for Heavy Ion Research,
             Oak Ridge National Laboratory,
             P. O. Box 2008, Oak Ridge, Tennessee 37831, 
             U.S.A.}
\affiliation{Institute of Nuclear Research of the 
Hungarian Academy of Sciences (Atomki) 
P.O.Box 51, H-4001, Debrecen, Hungary}
\affiliation{University of Debrecen, Faculty of Informatics, H-4001,Debrecen, 
P.O.Box 12,  Hungary}

\date{October 6 2005}
%
%
\begin{abstract}
The parameters of the nuclear liquid drop model, such as
the volume, surface, symmetry, and curvature constants, as well as bulk
radii, are extracted from
the non-relativistic and relativistic energy density functionals used in microscopic
calculations for finite nuclei. The microscopic liquid drop energy,
obtained self-consistently for a large sample of finite, spherical nuclei,
has been  expanded in terms of powers of $A^{-1/3}$
(or inverse nuclear radius) and the isospin excess (or neutron-to-proton asymmetry).
In order  to perform a reliable extrapolation in
the inverse radius, the calculations have been carried out for nuclei with 
huge numbers of nucleons, of the order of $10^6$.
The Coulomb interaction has been ignored to be able to approach nuclei of
arbitrary sizes and to avoid radial instabilities characteristic of systems
with very large atomic numbers. The main contribution to the fluctuating
part of the binding energy has been removed using  the Green's function method
to calculate the shell correction.
The limitations of applying the leptodermous expansion
to actual nuclei are discussed. While the leading  terms 
in the macroscopic energy expansion can be
extracted very precisely, the higher-order, isospin-dependent
terms are prone to large uncertainties due to finite-size effects.
\end{abstract}
\maketitle
%
%
\section{Introduction}

Bulk properties of atomic nuclei that depend in a smooth way
on the numbers of nucleons have been traditionally 
described in terms of macroscopic models, e.g.,
the liquid drop model (LDM) or the droplet model; for reviews, see 
Refs.\ \cite{[Mye77],[Mye82],[Has88],[Mol95],[Pom02]}. 
These phenomenological models, often augmented
by a  shell correction which is calculated using  average
single-particle potentials,  have been tuned up to
describe nuclear bulk properties to a  high precision. At the
same time, these models can be interpreted in the language of the
leptodermous expansion \cite{[Gra82]} that
sorts the various contributions to the  binding
energy of finite nuclei  in terms that have
transparent  physical meaning, e.g., volume, surface, symmetry, curvature,
and Coulomb energy. 

On the microscopic side,  self-consistent  mean-field models
employing density-dependent effective interactions or energy-density 
functionals are nowadays commonly used  in nuclear structure modeling.
The most prominent of these are the Skyrme-Hartree-Fock (SHF) method, 
the relativistic mean-field (RMF) approach (as well as their
Bogoliubov extensions),
and the Hartree-Fock-Bogoliubov method with the finite-range
Gogny force; for a recent review, see Ref.~\cite{[Ben03]}. These models
rely on  effective energy-density functionals with typically 6-10
parameters adjusted phenomenologically that  provide a global
description of all nuclei throughout the nuclear chart (with exception,
perhaps, of the lightest ones).  While the parameters of these
models can be organized and interpreted  using the
low-energy effective field theory of quantum chromodynamics 
\cite{[Fur97],[Fur97a],[Bur02]}, they do not
have  an immediate  interpretation in terms of the total number of nucleons
(or nuclear radius) and neutron excess. 
For that reason, it is both convenient and  practical \cite{[Ber05]}
to characterize nuclear energy-density functionals in terms of certain  
macroscopic parameters. The usual starting point is the  limit of the 
homogenous infinite nuclear matter, which is
simple to compute and which defines  the leading LDM characteristics, e.g., 
the volume energy, symmetry energy, or incompressibility. 

Nuclei have a  pronounced  surface; hence, a proper characterization of
surface properties is crucial. This task, however,
is far from easy. The usual means of characterizing surface properties
of the energy functional 
is through semi-infinite nuclear matter having a planar surface zone
(see Ref.~\cite{[Cen98a]} and references quoted therein). 
Early attempts
used semi-classical approaches to circumvent the enormous complexity of 
self-consistent calculations for the 
corresponding  mean field (see, e.g., Ref.~\cite{[Bra85]} for the
case of SHF). The limited  self-consistent calculations
in  SHF \cite{[Pea94],samyn} and RMF \cite{[Eif94a]} indicate  that the
quantum effects are non-negligible to the extent that they change the surface
parameters by about 5 $\%$. 

In  Ref.~\cite{[Dob02c]} the volume and surface contributions to the energy
density were extracted by assuming the Thomas-Fermi relation between
the local density and the kinetic energy density. The authors concluded that,
in the nuclear surface  zone, the gradient terms (absent in the homogenous
nuclear matter) are as important in defining the energy
relations as those depending on the local density. That is,
the nuclear surface cannot simply be regarded as a layer of nuclear matter at low
density. 

In this work, we extract the macroscopic LDM parameters by using a
large sample of finite, spherical nuclei, including huge systems
having $10^{5}$-$10^6$ nucleons. Based on the self-consistent SHF and
RMF results, we extract the macroscopic information from the large-scale 
trends by subtracting fluctuating shell corrections. The Coulomb
force is switched off to allow computation of very large nuclei. (We
thus concentrate on the strong component of the nuclear interaction.) This
way of analysis, using finite nuclei rather than semi-infinite matter,
is conceptually closer to existing nuclei, and it allows the
determination of curvature and surface symmetry effects. In essence,
the aim of this paper is two-fold.  First, by considering finite,
although huge, nuclei, we investigate the convergence of the
macroscopic expansion.  Second, by taking several SHF and RMF energy
functionals, we explore the relation between surface parameters and
the nuclear matter features of the underlying forces.

The paper is organized as follows. Section~\ref{macroscopic}
discusses the macroscopic energy expansions.  
The  details of our SHF and RMF models and shell-correction  calculations
are given in Sec.~\ref{models}. 
The extraction of LDM parameters is described in Sec.~\ref{LDM};
they are discussed in  Sec.~\ref{results}.
Finally, conclusions are drawn  in Sec.~\ref{conclusions}.

%
%
\section{The macroscopic energy}
\label{macroscopic}
According to the {\em Strutinsky Energy Theorem}
\cite{[Str67],[Str68],[Bra75],[Bra81]}, the energy per nucleon can be
decomposed into an average part (smoothly depending on the number of
nucleons) and the shell-correction term that fluctuates with particle
number reflecting the non-uniformities (bunchiness) of the
single-particle level distribution:
\begin{equation}\label{eqetot}
  \frac{E}{A}
  =
  {\cal E}
  =
  {\cal E}^{\rm(smooth)}+\delta{\cal E}^{\rm(shell)}.
\end{equation}
Macroscopic models, such as LDM,
deal with the smooth part. Consequently, in the following
we  concentrate on the average part of the binding energy:
\begin{equation}
  {\cal E}^{\rm(smooth)}
  =
  \frac{E}{A} -
  \delta{\cal E}^{\rm(shell)}
  \quad.
\label{eqetot1}
\end{equation}
The macroscopic energy can be parametrized in many ways. However, by far
the most successful macroscopic mass expressions are those rooted in 
the liquid-drop model (LDM) and in the   droplet model. They are respectively
outlined in Sec.~\ref{sec:LDM} and  Sec.~\ref{sec:droplet} below.

\subsection{Liquid-drop model}
\label{sec:LDM}

The LDM parameterizes the  binding energy of the nucleus ($Z$, $N$)
in equilibrium. Instead of proton and neutron numbers, it is convenient
to express the LDM energy through the mass number
and neutron  excess:
\begin{equation}
  A=N+Z 
  \quad,\quad
  I=\frac{N-Z}{N+Z}
  \quad.
\end{equation}
The macroscopic binding energy per nucleon can be expanded as
\begin{equation}
\begin{array}{rclclcl}
\displaystyle
{\cal E}^{\rm(LDM)}
& = & \multicolumn{5}{l}{{\cal E}^{\rm(smooth)}(A,I)}\\
& = & a_{\rm vol} & + & a_{\rm surf}A^{-1/3} & + & a_{\rm curv}A^{-2/3} \\[4pt]
&   &             & + & a_{\rm sym}I^2       & + & a_{\rm ssym}I^2 A^{-1/3} \\[5pt]
&   &             &   &                      & + & a_{\rm sym}^{(2)}I^4 \quad.
\end{array}
\label{eq:LDMansatz1}
\end{equation}
All the terms in Eq.~(\ref{eq:LDMansatz1}) have an immediate physical
interpretation. The bulk energy is given by the volume energy $a_{\rm vol}$, 
and changes with the neutron excess are accounted for by the
symmetry-energy term $a_{\rm sym} I^2$ and by the second-order
symmetry-energy term $a_{\rm sym}^{(2)} I^4$.  The most important
finite-size correction is the surface energy $a_{\rm surf}A^{-1/3}$,
followed by more subtle trends in terms of the curvature energy
$a_{\rm curv}A^{-2/3}$ and the surface-symmetry energy 
$a_{\rm ssym} I^2 A^{-1/3}$.

The sorting in columns indicates the level of importance of the terms.
Two different sorting criteria are used simultaneously: an expansion
of finite size effects (=surface effects) in terms of powers of
$A^{-1/3}$ (proportional to inverse radii) and, parallel to it, an
expansion in terms of the neutron-to-proton asymmetry $I^2$. The
second-order symmetry energy term $\propto I^4$ is not always included
in the macroscopic binding energy expression.  It has been considered,
e.g., in the context of the  Thomas-Fermi model \cite{[Mye98b]} and in a
discussion of strongly asymmetric matter within the RMF
\cite{[Liu02a]}.  We find that such a term appears naturally in the
hierarchy of Eq.~(\ref{eq:LDMansatz1}), and we shall demonstrate that
it is naturally present in the microscopic LDM expression.

At this point, it is worth noting that the shell energy per nucleon,
 $\delta{\cal E}^{\rm(shell)}$, scales with mass as $A^{-2/3}$
 \cite{[Boh75]}, i.e., it
 has the same dependence on the nuclear radius as the curvature term. 
Consequently, uncertainties associated with the extraction of   shell
corrections from self-consistent results,
and the presence of higher-order fluctuating terms that are not
accounted for by the Strutinsky procedure, can seriously
impact the values of higher-order terms in the leptodermous expansion.
We will see it very clearly in the results presented in Sec.~\ref{LDM}.
The ansatz (\ref{eq:LDMansatz1}) does not include explicit information about
the nuclear radius.  In fact, the LDM tacitly assumes a fixed radius
\begin{equation}
  R_0
  = 
  r_s A^{1/3}
\label{eq:rad0}
\end{equation}
where the Wigner-Seitz radius is typically
$r_s=$1.14-1.20 fm, which defines the saturation (equilibrium) density 
\begin{equation}
\rho_0
= \frac{3}{4\pi r_s^3}
\quad.
\end{equation}
A more general ansatz which allows a determination of the radius is 
provided by the droplet model 
presented below.

%
%
%
\subsection{Droplet model}
\label{sec:droplet}
The droplet model \cite{[Mye74]} (see, e.g., Ref.~\cite{[Mol95]} for a
recent implementation) includes the effect of the neutron skin and
non-uniformities in the nuclear density.  The two crucial parameters
of the droplet model that describe deviations from the equilibrium are
the neutron skin thickness $d$ and $\epsilon$, the relative deviation
in the bulk of the density $\rho$ from its nuclear matter value
$\rho_0$:
\begin{subequations} 
\begin{eqnarray}
  d 
  & = & 
  R_n - R_p 
  \quad, 
\\
  {\epsilon}
  & = & 
  \frac{R-R_0}{R}
  =
  - \frac{\rho-\rho_0}{3\rho_0},
\end{eqnarray}
\end{subequations}
where $R_0$ is the equilibrium radius of the droplet, see
eq.~(\ref{eq:rad0}).  Energy changes with $d$ and $\epsilon$ are
considered explicitly. For instance, the volume term is augmented by a
compression effect,
\begin{equation}
  a_{\rm vol}
  \longrightarrow
  a_{\rm vol}+\frac{1}{2}K\epsilon^2
  \quad,
\end{equation}
where the nuclear incompressibility coefficient is 
\begin{equation}
K \equiv 9 \rho^2_0 \frac{d^2}{d\rho^2} \frac{E}{A}\bigg|_{\rho = \rho_0}.
\end{equation}

The droplet model binding energy per particle is
considered a function of $A$, $I$, ${\epsilon}$, and $d$:
\begin{widetext}
\begin{equation}
\begin{array}{rclclcl}
  {\cal E}^{\rm(drop)}
  & = & \multicolumn{5}{l}{{\cal E}^{\rm(drop)}(A,I,\epsilon,d)}\\
  & = & 
  a_{\rm vol} 
  & + & 
  a_{\rm surf}A^{-1/3} 
  & + & 
  \tilde{a}_{\rm curv}A^{-2/3}
  + 2a_{\rm surf}A^{-1/3}{\epsilon} + \frac{K}{2}{\epsilon}^2
\\
  &&& + &
  a_{\rm sym}{I^2}
  &+&
  \tilde{a}_{\rm ssym}A^{-1/3}f(I,d)
  -
  3 a'_{\rm sym}{I^2} {\epsilon}
 +
  \tilde{a}^{(2)}_{\rm sym}{I^4}
\end{array}
\label{eq:dropansatz3}
\end{equation}
\end{widetext}
where $a'_{\rm sym}$ is defined as
\begin{equation}
a'_{\rm sym}
= \frac{\partial a_{\rm sym}}{\partial \rho} \bigg|_{\rho = \rho_0}
.
\end{equation}
The  function $f(I,d)$ is assumed to be   quadratic  in $I$ and $d$; it 
is determined by minimizing  the energy  with respect to $d$, see below.

%
%
\subsection{Relation  between droplet model and  LDM expansions}
\label{sec:more}

In this section, we recall the relation between 
the  LDM  expression
 (\ref{eq:LDMansatz1}) and the more detailed droplet mass formula
(\ref{eq:dropansatz3}).  The form
(\ref{eq:LDMansatz1}) is valid at the energy in equilibrium (i.e., where
radius and/or density have been adjusted to minimize the energy for a
given nucleus) while equation
(\ref{eq:dropansatz3}) allows  for separate tuning of 
 ${\epsilon}$ and $d$. The equilibrium energy ${\cal E}^{\rm(red)}(A,I)$
 is obtained
 by minimizing   
${\cal E}^{\rm(drop)}(A,I,{\epsilon},d)$  with respect to
${\epsilon}$ and $d$. 
At the equilibrium, $d= b I$, where $b$$\approx$1.4\, fm.
For the radial expansion ${\epsilon}$, one obtains
\begin{equation}
  {\epsilon}
  = 
  \frac{-2a_{\rm surf} A^{-1/3} + 3 a'_{\rm sym}I^2} {K}
  \quad.
\label{eq:trendrad}
\end{equation}
By substituting (\ref{eq:trendrad}) in (\ref{eq:dropansatz3}), one arrives at the 
LDM expression 
(\ref{eq:LDMansatz1})
where the leading parameters $a_{\rm vol}$, $a_{\rm surf}$, and 
$a_{\rm sym}$ remain unchanged while  the higher-order parameters
are redefined as:
\begin{subequations}
\label{eq:relateLDM}
\begin{eqnarray}
  a_{\rm curv}
  &=&
  \tilde{a}_{\rm curv} - 2a_{\rm surf}^2 \frac{\rho_0^2}{K}
  \quad,
\label{eq:intrcurv}\\
  a_{\rm ssym}
  &=&
  \tilde{a}_{\rm ssym} - 6a_{\rm surf}a'_{\rm sym}\frac{\rho_0^2}{K}
  \quad,
\label{eq:surfsym2}\\
  a^{(2)}_{\rm sym}
  &=&
  \tilde{a}^{(2)}_{\rm sym}-\frac{9}{2}(a'_{\rm sym})^2\frac{\rho_0^2}{K}
  \quad.
\label{eq:intrasmy2}
\end{eqnarray}
\end{subequations}
It is seen that while the leading parameters,
 $a_{\rm vol}$, $a_{\rm surf}$, 
    and $a_{\rm sym}$,  are defined unambiguously
in both LDM and droplet model expressions, the higher-order terms
differ. We shall determine the parameters of the leptodermous
expansion from the calculated ground state binding energies. The
corresponding mean-field configurations are stable points; hence, the
equilibrium philosophy of the LDM should apply. To deduce the droplet
model parameters $\tilde{a}_{\rm curv}$, $\tilde{a}_{\rm ssym}$, and
$\tilde{a}^{(2)}_{\rm sym}$, one should use the relations
(\ref{eq:relateLDM}).

\begin{table*}
\caption{\label{tab:coeffinout} 
LDM coefficients (in MeV) for the self-consistent mean-field
models applied in this work. The first block
(columns 2 and 3)  shows, for reasons of completeness, bulk
droplet-model parameters $\rho_0$ and $K$.  The second block 
(columns 4-6) shows the bulk parameters of the LDM
as computed in the limit of homogenous
 nuclear matter. 
The third block (semi-bulk; column 7) shows surface parameters from
semi-infinite nuclear matter calculations,  where available 
(Ref. \cite{samyn} for SHF and Ref.~ \cite{[Eif94b]} for RMF).  
The results for the parameters deduced from finite
nuclei as described below are shown in columns 8-10. 
The theoretical uncertainty on
surface energies is  0.05 MeV for SHF and 0.1 MeV for RMF. The curvature energies
are reliable within 0.5 MeV. The surface-symmetry energies have an uncertainty of
about 2 MeV for SHF and 10 MeV for RMF.  The lowest rows show the LDM  coefficients
adjusted  to data on finite nuclei.
The corresponding parameters of the droplet model (\ref{eq:dropansatz3})
can be deduced from the given LDM parameters through the relations
(\ref{eq:relateLDM}) plus the empirically supported
assumption (\ref{eq:hypo1}).
}
\begin{center}
\begin{ruledtabular}
\begin{tabular}{lccccccccc}
\noalign{\smallskip}
  & \multicolumn{5}{c}{bulk properties } & semi-bulk
  & \multicolumn{3}{c}{from  finite nuclei} \\
\noalign{\smallskip}\cline{2-6}\cline{7-7}\cline{8-10}\noalign{\smallskip}
 Model & $\rho_0$  & $K$ & $a_{\rm vol}$ & $a_{\rm sym}$ &  $a'_{\rm sym}$ & 
         $a_{\rm surf}^{\text{(NM)}}$ &  $a_{\rm surf}$ & 
         ${a}_{\rm curv}$ &  ${a}_{\rm ssym}$ \\
\noalign{\smallskip}\hline\noalign{\smallskip}
 SkM$^*$&0.1603& 216.6 & -15.752 &  30.04 & 95.25 & 17.70 & 17.6 & 9 & -52 \\
 SkP  & 0.1625 & 201.0 & -15.930 &  30.01 & 40.43 & 18.22 & 18.2 & 9.5 & -45 \\
 BSk1 & 0.1572 & 231.4 & -15.804 &  27.81 & 15.76 & 17.54 & 17.5 & 9.5 & -36 \\
 BSk6 & 0.1575 & 229.2 & -15.748 &  28.00 & 35.67 &       & 17.3 & 10 & -33 \\
 SLy4 & 0.1596 & 230.1 & -15.972 &  32.01 & 95.97 & & 18.4 & 9 &  -54 \\
 SLy6 & 0.1590 & 230.0 & -15.920 &  31.96 & 99.48 & 17.74 & 17.7 & 10 & -51\\
 SkI3 & 0.1577 & 258.1 & -15.962 &  34.84 & 212.47 & & 18.0 & 9 &  -75 \\
 SkI4 & 0.1601 & 247.9 & -15.925 &  29.51 & 125.80 & & 17.7 & 9 &  -34 \\
 SkO  & 0.1605 & 223.5 & -15.835 &  31.98 & 163.50 & & 17.3  & 9 & -58 \\
\noalign{\smallskip}\hline\noalign{\smallskip}
 NL1  & 0.1518 & 211.3& -16.425  & 43.48 & 311.18 &      & 18.8 &9  &-110\\
 NL3  & 0.1482 & 271.7& -16.242  & 37.40 & 269.16 & 18.5 & 18.6 &7 & -86\\
 NL-Z & 0.1509 & 173.0& -16.187  & 41.74 & 299.51 &      & 17.8 &9 & $<$\,-125\\
 NL-Z2& 0.1510 & 172.4& -16.067  & 39.03 & 281.40 & 17.7 & 17.4 &10 & -90 \\
\noalign{\smallskip}\hline\noalign{\smallskip}
 LDM \cite{[Mol95]}
      & 0.153  &       & -16.00   & 30.56 &   &   & 21.1 &  & -48.6    \\
 LDM \cite{[Pom02]} 
      & 0.1417 &       & -15.848  & 29.28 &   &   & 19.4 & &  -38.4    \\
 LSD \cite{[Pom02]}  
      & 0.1324 &       & -15.492  & 28.82 &   &   & 17.0  & 3.9 &   -38.9 \\
\noalign{\smallskip}
\end{tabular}
\end{ruledtabular}
\end{center}
\end{table*}

%
%
\section{Self-consistent Models}
\label{models}

We employ two variants of self-consistent mean-field models: SHF and
 RMF. They are explained in detail in Ref.~\cite{[Ben03]}. Both
approaches provide a functional form for the energy density with a
good handful of free parameters. These have been adjusted to
phenomenological data by different groups and with different
bias. Thus there exist various parametrizations on
the market which provide a fairly good description of basic
nuclear bulk properties in the valley of stability, but differ in
other aspects as, e.g., excitations, fission barriers, neutron matter
properties, or electromagnetic form factors.

In this work,
we have chosen a small subset of Skyrme forces
which perform well  for  the basic ground-state properties 
and have
sufficiently different properties which allows one to explore the
possible variations among parameterizations.  This subset
contains: SkM$^*$ \cite{[Bar82]}, 
SkP \cite{[Dob84]}, SLy4
\cite{[Cha95a]}, 
SLy6 \cite{[Cha98]}, SkO \cite{[Rei99]},
BSk1 \cite{[Sam02]},
BSk6 \cite{[Gor03]},
and SkI1, SkI3, and SkI4 from
Ref.~\cite{[Rei95]}. 
SkP, SkO, and BSk1 have effective nucleon
mass around one, leading to a comparatively large density
of single-particle levels. All other SHF forces employed here
have smaller effective masses.
Interesting here is the double BSk1 with BSk6. Both forces
were fitted using a similar strategy and data pool, but
have different effective mass $m^*/m=1.05$ for BSk1 and $0.8$ for
BSk6. The force
SLy6  was adjusted with particular emphasis on isotopic
trends and neutron matter.
The functionals SkI3, SkI4, and SkO have a generalized isovector
spin-orbit interaction compared to all other forces. Some forces,
i.e.\ SkP, BSk1, and BSk6, include a $\tensor{J}^2$ term 
(where $\tensor{J}$ is the spin-orbit tensor) with a coupling 
constant related to the surface and effective mass terms, while all
others do not. All the selected forces perform reasonably well 
concerning the total energy and radii for nuclei close to the 
valley of stability, with some different bias on particular 
observables. In particular, BSk1 and BSk6 are fits to all available
masses, but only masses.

As seen in Table \ref{tab:coeffinout}, the values for the volume 
energy coefficient $a_{\text{vol}}$, saturation density $\rho_0$,
incompressibity $K$, and surface energy coefficient $a_{\text{surf}}$
are quite similar, with slight systematic differences between SHF
and RMF that already have been noticed earlier; see \cite{[Ben03]}
and references given therein. Obvious large variations occur for 
properties which are not fixed precisely by nuclear matter and 
ground-state characteristics.

As in  SHF, there exist many RMF parameterizations  which differ in details. 
For the purpose of  the present study, we choose 
the most successful (or most commonly used)  ones: 
NL1 \cite{[Rei86]}, NL-Z \cite{[Ruf88a]}, NL-Z2 \cite{[Ben99a]}, 
and  NL3 \cite{[Lal97]}. The parameterization NL1 is a fit of the RMF 
along the strategy of Ref.~\cite{[Fri86]}. The NL-Z parametrization
is a refit of NL1 where the correction for spurious 
center-of-mass motion is calculated from the actual many-body
wave function, while
NL-Z2 is a recent variant of NL-Z with an improved isospin dependence.
The force NL3 stems from a fit including exotic 
nuclei, neutron radii, and information on giant resonances.
All the above  parameterizations provide a good description of 
binding energies, charge radii, and surface thicknesses 
of stable spherical nuclei with the same overall quality as 
the SHF model. As seen in Table \ref{tab:coeffinout}, however,
the nuclear matter properties of the RMF forces 
show some systematic differences as compared to Skyrme forces. 
All RMF forces have comparable small effective
masses around $m^*/m \approx 0.6$. (Note that the effective mass in 
RMF depends on momentum; hence  the effective mass 
at the Fermi energy is approximately $10\%$ larger.) Compared with
SHF models, the absolute value of the energy per nucleon is 
systematically larger, with values around $-16.3$ MeV, 
while the saturation density is always 
slightly smaller with typical values around 0.15 nucleons/fm$^{3}$.
The incompressibility of the RMF forces ranges from low
values around 170 MeV for NL-Z to $K$=270 MeV for NL3.
There are also differences in isovector properties; the symmetry
energy coefficient of all RMF forces is systematically larger than for
SHF interactions, with values between 37.4 MeV for NL3 and
43.5 MeV for NL1.

The absolute variation of the LDM energy coefficients found in 
Table \ref{tab:coeffinout} should be put  into perspective. There is a 
notable difference between the variation of the coefficients in the
LDM expansion on the one hand and the actual variation of the energy 
on the other hand. Let us consider the heavy nucleus $^{250}$Fm with 
\mbox{$A=250$} and \mbox{$I=0.2$} as an example. 
The difference in $a_{\text{vol}}$ between SkM* and NL1 
is 0.67 MeV, which seems to be small. It leads, however, to an energy 
difference of about 160 MeV, which amounts to a significant 
fraction  ($\sim$10 $\%$) of the total binding energy of this nucleus.
The difference in the symmetry energy coefficient, $a_{\text{sym}}$
between the same two forces is 3.44 MeV and appears to be much more 
significant that the difference in $a_{\text{vol}}$. However,
 the $I^2$ factor suppresses the 
difference in the  symmetry energy between SkM* and NL1 to 34.4 MeV. 
Even the 60 MeV difference between the surface symmetry energy
coefficients of SkM* and NL1 only gives about 100 MeV difference in
binding energy between the two forces, which is comparable with, but 
still smaller than, the energy difference arising from the small 
difference in the volume energy coefficient.

It is interesting to check to what extent
the differences in $\rho_0$, $K$, and $a'_{\text{sym}}$
influence the macroscopic parameters and whether a simple correlation
between spectroscopy and macroscopy can be found. This will
be done in Sec.~\ref{sec:systema}, where  we will consider
a larger variety of parameterizations and perform 
dedicated variations of special features as, e.g., the effective nucleon
mass. 

According to Eq.~(\ref{eqetot1}),
the average   energy  is obtained by subtracting the
shell-correction energy from the self-consistent value.
The shell energy  is computed using the same prescription as
outlined in Refs.~\cite{[Kru00b],[Ver00]}.
This procedure is based on the Green's function approach to the level
density and the generalized plateau  condition of  Ref. \cite{[Ver98]}.
The advantage of this procedure is that it properly takes care of the  
continuum positive-energy states which unavoidably come into
play in the self-consistent approach. 
In our calculations, we include a large space 
of single-particle states up to 60 MeV  above the Fermi energy.
Since most of these states are continuum states,
the contribution from a particle gas (treated in the same numerical
box) has to be removed. To meet the generalized plateau  condition,
it is assumed that the deviation of the smoothed level density
from linearity is  minimal in a wide  energy interval [--50,--20] MeV.

Our calculations are restricted to spherical symmetry; they were carried
out using numerical techniques  described in detail in Ref.~\cite{[Rei89a]}. 
The Coulomb force is ignored to allow an
extension to nuclei of arbitrary sizes.  Pairing correlations
are  neglected as well. However, open-shell nuclei were treated in the filling
approximation in which we use a very small pairing force (factor of 10 
smaller than usual). The center-of-mass (c.m.)  correction is
included. We take care to use precisely the c.m.\  recipe that 
is attached  to a given force \cite{[Ben03]}. This is crucial because it 
is known that the actual form of the c.m.\ correction has 
a significant impact on the surface properties \cite{[Ben00d]}.

%
%
\section{Extraction of LDM parameters}
\label{LDM}
\subsection{Bulk parameters}
\label{sec:bulk}

The bulk parameters in the leptodermous expansion 
are those proportional to $A^0$. They represent terms which do not
vanish in the \mbox{$A \to \infty$} limit. In the LDM ansatz 
(\ref{eq:LDMansatz1}) the bulk parameters  are: the volume energy constant  
$a_{\rm vol}$, the symmetry energy constant  $a_{\rm sym}$, and the 
second-order symmetry energy parameter ${a}_{\rm sym}^{(2)}$. 
The droplet model (\ref{eq:dropansatz3}) additionally introduces
the incompressibility $K$, the density-slope of the symmetry
energy $a'_{\rm sym}$, and the equilibrium density $\rho_0$. All these
parameters can easily be computed in the limit
of the homogenous bulk nuclear matter, see, e.g., Ref.~\cite{[Ben03]}.

Asymmetric nuclear matter shows an interesting trend, which sheds some light on
the relation between the LDM and droplet model. The bulk part of the 
LDM energy can be written as:
\begin{equation}
\label{eq:reducedbulk}
  {\cal E}^{\rm(LDM)}
  =
  a_{\rm vol} 
  +
  a_{\rm sym} I^2
  +
  a_{\rm sym}^{(2)}I^4
  .
\end{equation}
In order to concentrate on the isospin  dependence,
we introduce an effective symmetry energy parameter  as
\begin{eqnarray}
\label{eq:effbulk}
  a_{\rm sym}^{\rm eff}
  &=&
  \frac{{\cal E}(I)-{\cal E}(I\!=\!0)}{I^2}
\\  
  &=&
  a_{\rm sym}+a_{\rm sym}^{(2)}I^2
  \quad.
\nonumber
\end{eqnarray}
The first line serves as a general definition. The second line then is specific
to the bulk limit.

\begin{figure}
\centerline{\epsfig{figure=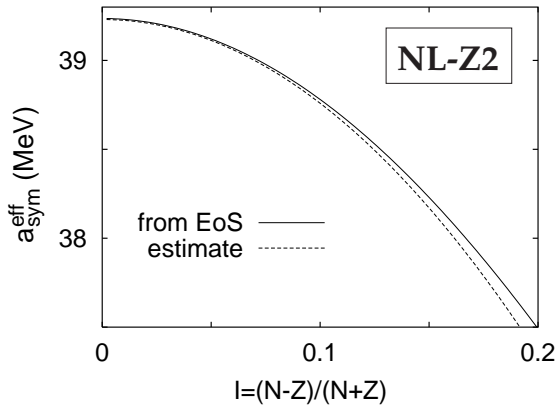,width=8.2cm}}
\caption{\label{fig:effasy_nlz2}
Solid line: the effective symmetry energy parameter (\ref{eq:effbulk}) in
asymmetric nuclear matter computed with NL-Z2 as a function of 
  neutron excess  $I$. Dotted line: 
the trend (\ref{eq:predI2})
using the bulk parameters of NL-Z2, see Table
\ref{tab:coeffinout}. 
 }
\end{figure}

Figure \ref{fig:effasy_nlz2} shows the effective symmetry energy
parameter  obtained in the NL-Z2 model. Let us now recall that
$a_{\rm sym}^{(2)}$ is related to the droplet parameter
$\tilde{a}_{\rm sym}^{(2)}$ through Eq. (\ref{eq:intrasmy2}).
This suggests that there is no explicit second-order isospin correction 
to the symmetry energy in the droplet model:
\begin{subequations}
\label{eq:hypo}
\begin{equation}
  \tilde{a}_{\rm sym}^{(2)}=0,
\label{eq:hypo1}
\end{equation}
which yields
\begin{equation}
  a_{\rm sym}^{\rm eff}(A\!=\!\infty,I)
  =
  a_{\rm sym}
  -
  \frac{9}{2}(a'_{\rm sym})^2\frac{\rho_0^2}{K}I^2
  \quad.
\label{eq:predI2}
\end{equation}
\end{subequations}
It is seen in Fig.~\ref{fig:effasy_nlz2}  that the estimate
(\ref{eq:predI2}) matches the exact result extremely well.
The deviations are  small and predominantly
$\propto I^4$, thus going beyond the present expansion.
(Higher-order isospin corrections were in fact considered in 
Ref.~\cite{[Liu02a]} in the context of RMF and infinite nuclear matter.) 
We have checked that  the assumption (\ref{eq:hypo}) is well fulfilled 
for all the RMF and SHF functionals  used in this work. The isospin 
dependence of  $a_{\rm sym}^{\rm eff}(A\!=\!\infty,I)$ in SHF is much 
weaker than in RMF. This is consistent with Eq.~(\ref{eq:predI2}) and 
droplet-model parameters displayed in Table \ref{tab:coeffinout}.
Finally, corroborating evidence of a very small $\tilde{a}_{\rm sym}$  
term  comes from Brueckner-Hartree-Fock studies of asymmetric matter based on 
realistic nucleon-nucleon and three-body forces \cite{[Bor98],[Zuo99],[Zuo02]}.

\subsection{Isospin-independent surface parameters}
\label{sec:isoscalar}

The leading isospin-independent parameters characterizing finite-size
(surface) terms in the LDM are the surface and curvature energy 
coefficients.
We deduce them from the systematics of binding energies of spherical
nuclei.

\begin{figure}
\centerline{\epsfig{figure=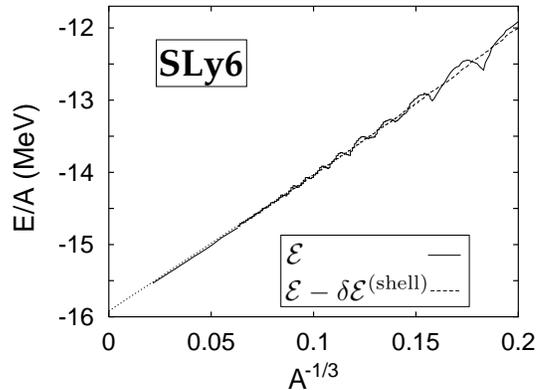,width=8.2cm}}
\caption{\label{fig:testhell}
The binding energy per nucleon  ${\cal E}(A,I)$ for isospin-symmetric ($I=0$) 
nuclei computed with SLy6 as a function  $A^{-1/3}$ before (solid line)
and after (dotted line) subtracting the shell correction 
$\delta{\cal E}^{\rm(shell)}$.
The fine dotted line connects the last value  from finite nuclei with the
nuclear matter limit.
 }
\end{figure}

Figure \ref{fig:testhell} shows the systematics of binding energies per
nucleon  predicted by SLy6.  The smooth component obtained by
subtracting $\delta{\cal E}^{\rm(shell)}$ is also indicated. At very
large values of $A$ (i.e. small $A^{-1/3}$), 
the binding energy per nucleon nicely converges to
a straight line which demonstrates the validity of the hierarchical LDM
(or droplet) ansatz and, not surprisingly, hints at the dominating role
played by the surface energy as first leading correction to the bulk.
The plot  also illustrates the effect of quantum shell fluctuations. The
actual binding energy oscillates around the average trend; the amplitude
of shell oscillations increases in lighter nuclei, consistent with the
expected $A^{-2/3}$-dependence discussed earlier in  Sec.~\ref{sec:LDM}. By
subtracting the  shell correction, one obtains a fairly smooth trend, at
least at this level of analysis.

\begin{figure}
\centerline{\epsfig{figure=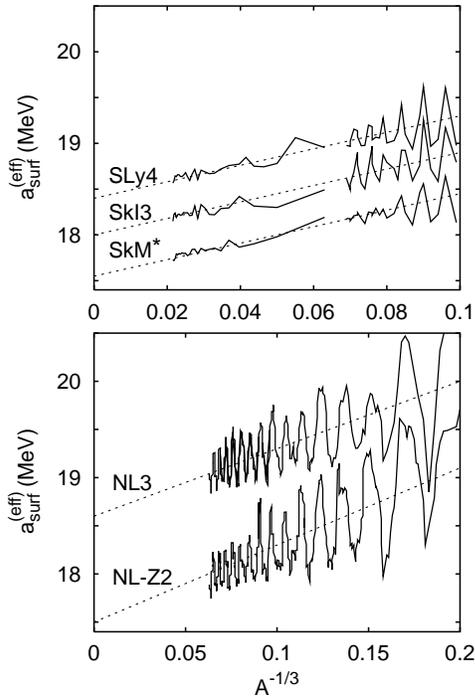,width=8.2cm}}
\caption{\label{fig:asurf} 
Effective surface-energy coefficient (\ref{eq:relatecurve})
versus $A^{-1/3}$ for several SHF (top) and RMF (bottom) functionals, as
indicated. A straight-line fit to the data is marked by a dashed line. 
}
\end{figure}

To extract the surface- and curvature-energy coefficients, it is convenient
to introduce the effective surface-energy coefficient:
\begin{subequations}
\begin{eqnarray}
\label{eq:relatecurve}
  a_{\rm surf}^{\rm(eff)}
  &=&
  \big[ {\cal E}^{\rm(smooth)}(A,0)-a_{\rm vol}\big] A^{1/3}
\\
  & =  & 
  a_{\rm surf} + {a}_{\rm curv}A^{-1/3} \label{asurrf}
  \quad,
\end{eqnarray}
\end{subequations}
which is  a function of system size $A$.  The surface-energy coefficient
$a_{\rm surf}$ is obtained by extrapolating  $A^{-1/3}\longrightarrow
0$. The curvature ${a}_{\rm curv}$ is then obtained from the slope of 
(\ref{asurrf}).

Figure \ref{fig:asurf} shows the effective surface energies.
Note that these are drawn from smoothed energies (i.e. after subtraction of
the shell corrections). The remaining fluctuations are due to
higher-order shell effects \cite{[Bra75],[Bra81]} which cannot
be accounted for by the generalized Strutinsky procedure. 
The construction
(\ref{eq:relatecurve}) of the effective surface energy amplifies those 
residual fluctuations dramatically; it is only by virtue of the smoothed
energy that one can see any clear trend. Thus atop these remaining shell
fluctuations, one can recognize a definite  slope (note
the very narrow energy window), which is related to the curvature
energy. By performing a straight-line fit to the data, surface- and
curvature-energy coefficients can be extracted.

Note that different $A^{-1/3}$ scales are used for SHF and RMF.  The
reason is that in SHF cases we have been able to carry up calculations
for really huge nuclei (\mbox{$A=10^6$}). This extends the scale to smaller
$A^{-1/3}$ and allows us to ignore some data points for lighter systems
where residual shell fluctuations are large. Unfortunately, we were
unable to approach similarly large nuclei in RMF; hence, we had to
scale up to \mbox{$A^{-1/3}=0.2$} to get sufficient data for
extrapolation. In any case, one sees that one can extract reliably
well surface and curvature energies from the trends displayed in
Fig. \ref{fig:asurf}.  We estimate an uncertainty in $a_{\rm surf}$ to
be about 0.05 MeV for SHF and 0.1 MeV for RMF. The curvature
coefficient is determined within about $\pm 0.5$ MeV.

\subsection{Surface-symmetry coefficient}

\begin{figure*}
\centerline{\epsfig{figure=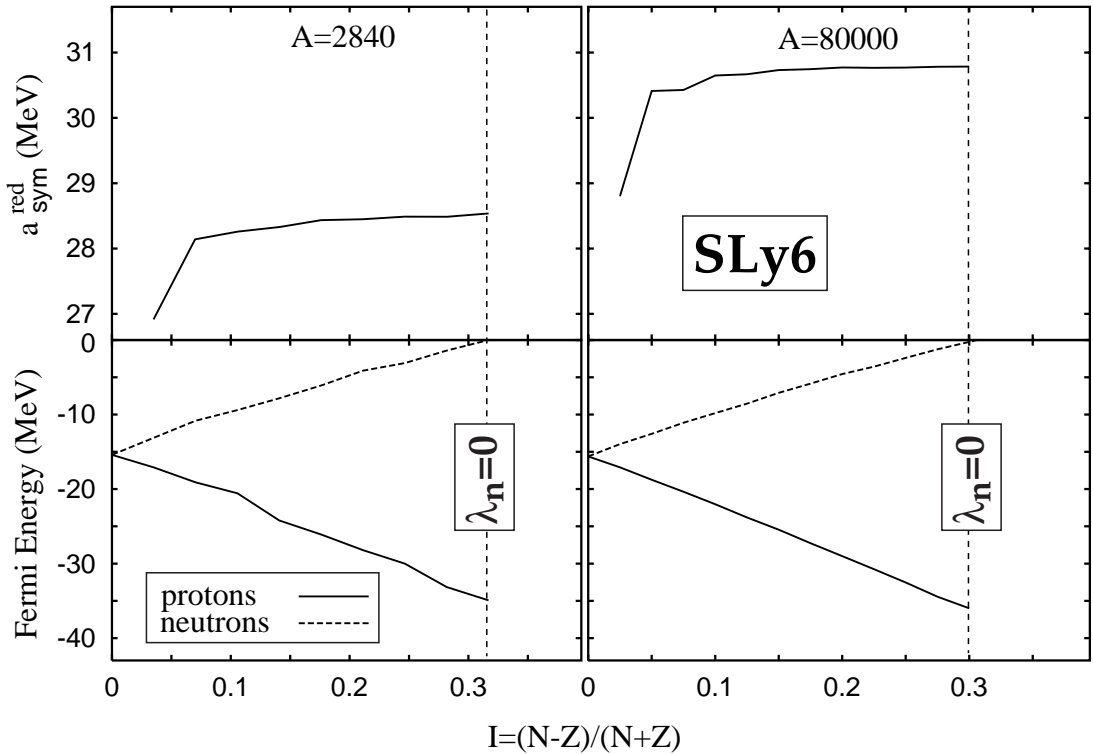,width=16.0cm}}
\caption{\label{fig:eferm}
Top: reduced effective symmetry-energy coefficient (\ref{eq:effasyred}) 
in SLy6 versus $I$ for two system sizes ($A$=2840, left; $A$=80,000, right).
Bottom: corresponding proton and neutron Fermi energies.
 }
\end{figure*}

In order to deduce the isospin-dependent 
surface-symmetry coefficient,  we come back to  
the effective symmetry parameter  (\ref{eq:effbulk}) now considering finite 
$A$ and identifying ${\cal E}\leftrightarrow{\cal E}^{\rm(smooth)}$.
As it is convenient to subtract the known  infinite-matter 
trend of Eq.~(\ref{eq:predI2}), we introduce  the reduced effective
symmetry-energy  coefficient:
\begin{widetext}
\begin{eqnarray}
  a_{\rm sym}^{\rm red}(A,I)
  &=& \frac{1}{I^2}
  \left({\cal E}^{\rm(smooth)}(A,I)-{\cal E}^{\rm(smooth)}(A,0)
   +\frac{9(a'_{\rm sym})^2\rho_0^2}{2K}I^4\right)
  \quad.
\label{eq:effasyred}
\end{eqnarray}
\end{widetext}
Figure \ref{fig:eferm} demonstrates how such methodology works. The
symmetry energy involves a difference of smoothed energies; hence, a
difference of shell corrections. These corrections are prone to 
uncertainties, as we have already seen  in Sec.~\ref{sec:isoscalar}.
At small values of $I^2$,  the remaining uncontrolled energy
fluctuations are amplified in the finite difference (\ref{eq:effasyred}), 
and the result is not reliable. Fortunately, at larger values of 
\mbox{$I\geq 0.1$}, where the isospin-dependent terms
dominate over remaining shell fluctuations, one always obtains a stable and 
well-defined  plateau. The value of $I$ is limited from above
by the neutron drip line. Indeed,  
at  \mbox{$I \approx 0.3$}  the neutron Fermi energy becomes positive 
and the self-consistent solution can  no longer be trusted.
Therefore, for further analysis, we introduce
an  $I$-averaged reduced effective symmetry-energy coefficient:
\begin{equation}
  \overline{a_{\rm sym}^{\rm red}}
  =
   {\int_{0.1}^{0.2}dI \, a_{\rm sym}^{\rm red}(A,I)}\;\left/\;
       {\int_{0.1}^{0.2}dI}\right.
  \quad.
\label{eq:aversym}
\end{equation}
The surface symmetry energy coefficient is obtained by plotting
$\overline{a_{\rm sym}^{\rm red}}$ versus $A^{-1/3}$. The slope for small
$A^{-1/3}$  corresponds to ${a}_{\rm ssym}$, similarly as it was
done in Sec.~\ref{sec:isoscalar} for deducing the surface and curvature
parameters.  An effective surface-symmetry constant can thus be obtained
from
\begin{equation}
  {a}_{\rm ssym}^{\rm(eff)}
  =
  \left(\overline{a_{\rm sym}^{\rm red}}-a_{\rm sym}\right)A^{1/3}
  \quad.
\label{eq:surfsymaver}
\end{equation}

\begin{figure}
\centerline{\epsfig{figure=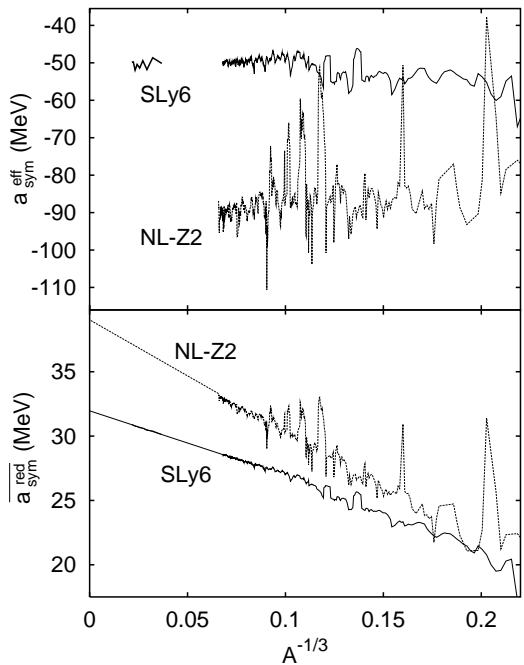,width=8.2cm}}
\caption{\label{fig:aversym_bw}
Bottom: the averaged reduced symmetry-energy coefficient (\ref{eq:aversym})
in  SLy6  and NL-Z2;
Top: the corresponding  effective surface-symmetry coefficient 
(\ref{eq:surfsymaver}).
 }
\end{figure}

Figure \ref{fig:aversym_bw} shows typical results of such
analysis. It is gratifying to see that
the effective symmetry-energy  coefficients (\ref{eq:aversym}) 
are consistent with the corresponding bulk values when extrapolating back
to \mbox{$A^{-1/3} \to 0$}. The quality of the deduced
slope (= surface asymmetry) can be assessed by inspecting the rescaled
quantity (\ref{eq:surfsymaver}) shown in the upper panel. 
Clearly, very large $A$, i.e., very small $A^{-1/3}$, are
required to see convergence to a strictly horizontal trend. With the
present data set, one may attach $\sim$10 $\%$ relative uncertainty 
(or about 2 MeV absolute error) to  ${a}_{\rm ssym}$.

%
%
%
\subsection{Radii}
\begin{figure}
\centerline{\epsfig{figure=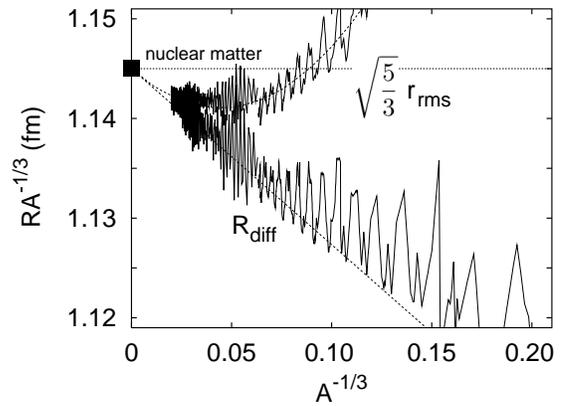,width=8.0cm}}
\caption{\label{fig:nucmatlimit}
Nuclear radii (multiplied by $A^{-1/3}$ to remove the overall mass dependence)
for $N=Z$ nuclei computed with SLy6 and (partly) corrected for shell 
fluctuations using the Strutinsky method. The lower  curve shows the  
diffraction radii and the upper curve shows the  r.m.s.\ radii 
$\sqrt{5/3}r_{\rm rms}$. The nuclear matter Wigner-Seitz
value of 1.145 fm is indicated by a box and a horizontal dashed  line. 
The droplet model estimate (\ref{dropletradius}) of the diffraction radius 
and the estimate (\ref{Rdiff})  of  the r.m.s. radius are shown 
(dotted lines). It is gratifying to see the agreement between  the 
droplet-model estimates and the results of self-consistent calculations 
for finite nuclei. 
}
\end{figure}

By employing the equilibrium value of $\epsilon$ (\ref{eq:trendrad}), we can 
estimate the droplet-model radius as
\begin{equation}\label{dropletradius}
R=R_0(1-{\epsilon}).
\end{equation}
It is worth checking the performance of that recipe. To this end,
we have extracted diffraction radii and r.m.s.\ radii from the SLy6 
calculations of large (and huge) spherical nuclei with $N$=$Z$. 
Figure~\ref{fig:nucmatlimit} displays the  nuclear radii corrected for 
shell fluctuations up to $A$=4000. It is seen that the estimate 
(\ref{dropletradius}) evaluated with the SLy6 parameters of 
Table \ref{tab:coeffinout} nicely approximates the actual results. 
The r.m.s. radii are systematically larger than the diffraction radii  
which is no surprise because in the  Helm model both are  related via
\begin{equation}
r_{\rm rms}
= \sqrt{\frac{3}{5}}\sqrt{R_{\rm diff}^2+5\sigma^2}
\end{equation}\label{Rdiff}
where  $\sigma$ is the surface thickness coefficient (related to the first 
maximum of the form factor) with a typical value of 1\,fm.
The estimate (\ref{Rdiff}) is indicated in Fig. \ref{fig:nucmatlimit},
and it is shown to be fully consistent with the self-consistent results for 
the r.m.s. radii. Moreover, our analysis makes it very clear that the
diffraction radius (or box-equivalent) is indeed the appropriate quantity 
entering the droplet model.

%
%
\section{Results and discussion}
\label{results}
%
%

\subsection{Existing energy functionals}\label{existingf}

Table \ref{tab:coeffinout} collects the calculated LDM parameters for
the SHF and RMF parametrizations introduced in Sec.~\ref{models}. The
bulk parameters are computed for the homogenous nuclear matter.  The
surface-related parameters are deduced from the present analysis of
finite nuclei. We also show for comparison the surface-energy
coefficients, $a_{\rm surf}^{\rm (NM)}$, obtained from quantum mechanical
calculations for the semi-infinite nuclear matter 
\cite{samyn,[Eif94b]}. The comparison of
$a_{\rm surf}^{\rm (NM)}$ with the values of $a_{\rm surf}$ deduced from
finite nuclei shows a nice agreement between both methods of analysis.

The surface-energy and curvature-energy coefficients are fairly robust
quantities in that the variations throughout all forces are small.
The curvature energy coefficient, $9.5\pm 0.5$ MeV, is smaller than
other LDM constants but certainly nonzero. Moreover, its variations
remain small throughout all forces ``from the shelves'', below the
 uncertainties of our analysis. In that context, we want to
comment on the investigations of Ref.~\cite{[Sto88b]} (see also
discussion in \cite{[Pom02]}) where it was
claimed that one needs small curvature-energy coefficients to obtain a good fit
to fission barriers.  While we find about the same curvature energy for a
variety of forces, the fission barriers of actinides and superheavy
elements  investigated in \cite{[Bur04]} turned out to be strongly 
force-dependent  for a similar pool of interactions. We suspect that 
another key parameter affecting  fission barriers lies in the
isovector channel. As one hint we will see in Sec.~\ref{sec:systema} 
that strong  variations of the symmetry-energy coefficient can 
influence $a_{\rm curv}$.

The surface energy is more important and its (small) variations
are larger than the uncertainty of about $\pm 0.1$ MeV.  The observed
trends can be sorted in various ways.  One important aspect is
that $a_{\rm surf}$ depends to some extent on the strategy used when
fitting a functional. It is obvious that the weight which was given to
light nuclei in the fit has an influence on the surface coefficients,
as well as whether surface properties were considered (e.g.,
the surface thickness in the case of SkI3 and SkI4).
Another crucial ingredient in the trends with $A^{-1/3}$ is the way in
which the center-of-mass (c.m.) correction was implemented
\cite{[Ben00d]}. In the sample of forces considered here, there are
two forces, SLy4 and SLy6, which were fitted with precisely the same
strategy but differ in their treatment of c.m. correction.  The
difference of 0.7 MeV in $a_{\rm surf}$ is quite remarkable and
represents mainly the difference between the recipes for the c.m.
correction used \cite{[Ben00d]}.  We think
that the recipe used in SLy6, namely to compute the c.m. energy from
the given mean field state, is better microscopically motivated. Thus
the lower value for $a_{\rm surf}$ is probably more realistic.
However, this has yet to be explored in more detail.

The situation for all the isospin-dependent LDM
parameters is quite different. That begins with
the large discrepancy between $a_{\rm sym}$-values  for RMF and
SHF. In fact, even within SHF alone there is a much larger variation in  the
symmetry energy than appears from  Table~\ref{tab:coeffinout}; see, e.g.,
the discussion in Ref.~\cite{[Rei99b]}. Extended  RMF functionals also show 
significant uncertainty in $a_{\rm sym}$ \cite{[Typ01]}. Even more 
pronounced are fluctuations in the surface-symmetry coefficient 
$a_{\rm ssym}$. By inspecting Table~\ref{tab:coeffinout}, one can see a 
rough correlation between $a_{\rm sym}$ and $a_{\rm ssym}$. This will be 
put on a firmer ground in Sec.~\ref{sec:systema} below where 
systematic variations of  functionals are discussed.

\subsection{Systematic variations of functionals} 
\label{sec:systema}

The discussion of Table \ref{tab:coeffinout} in Sec.~\ref{existingf}
has indicated several features which deserve closer inspection. To
this end, we perform systematic variations of 
key properties of the functional, as, e.g. effective nucleon mass
$m^*/m$ or symmetry energy $a_{\rm sym}$. The strategy is to
vary only one chosen property 
while keeping all other features fixed. A set of SHF forces has
been produced that way by fitting the parameters always to the same
set of data (energies, charge form factors, spin-orbit splitting)
while putting a constraint on the required additional feature. This
was done formerly for the purpose of studying trends in the giant
resonances \cite{[Rei99b]}. We consider these families of functionals
in this work to inspect trends and correlations in a systematic
manner.

\begin{figure*}
\centerline{\epsfig{figure=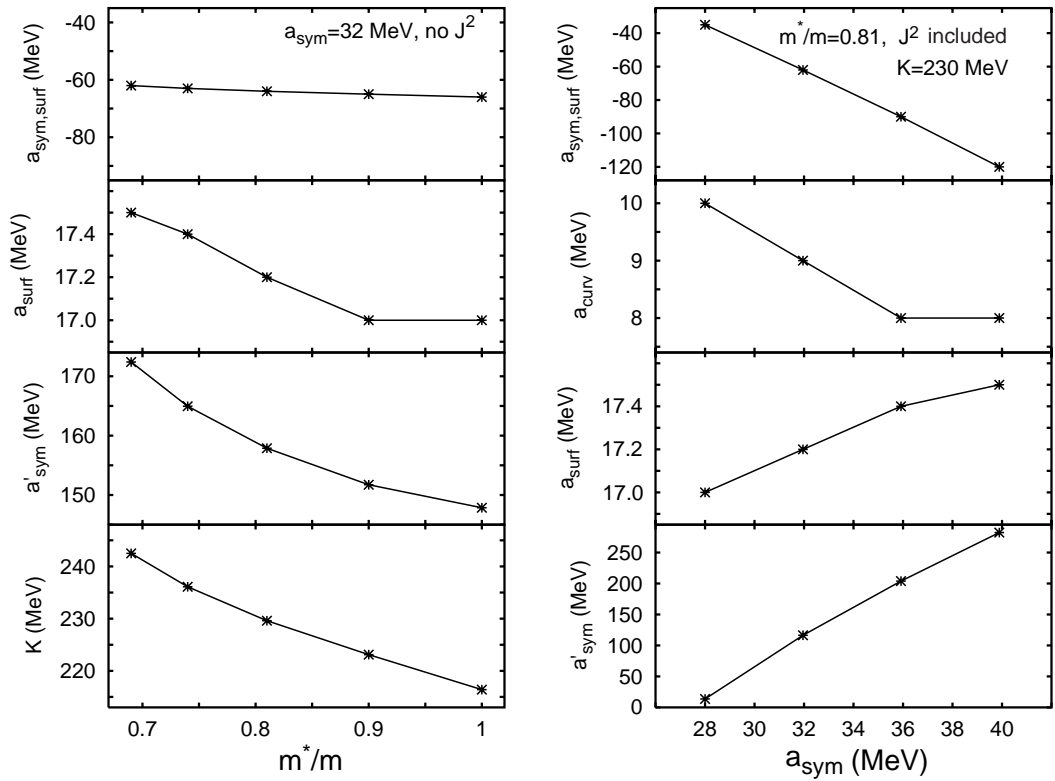,width=16cm}}
\caption{\label{fig:Tvary}
Dependence of  LDM  parameters on  effective mass (left) and the
symmetry-energy coefficient  (right). The incompressibility is not shown in 
the right panel because it does not change. The curvature energy coefficient
is not shown in the left panel because it varies very little as a function
of $m^*/m$. The extended Skyrme functionals of Ref.~\cite{[Rei99b]} are used.
}
\end{figure*}

Figure \ref{fig:Tvary} shows results of two sets of calculations. The
first group in the left panel contains functionals with systematically
varied effective mass $m^*/m$. The functionals belonging to the second
set, shown in the right panel, vary $a_{\rm sym}$.  
Looking at the left panel, we find sizable variations with effective
mass concerning $a_{\rm surf}$, $K$, and $a'_{\rm sym}$.  All those
parameters decrease with $m^*/m$.  The surface-symmetry coefficient
$a_{\rm ssym}$ is fairly insensitive to $m^*/m$. The curvature
energy coefficient ${a}_{\rm curv}$ (not shown here) varies only between 9
and 10 MeV which is very little in view of the theoretical uncertainty
in this parameter.
Different trends are seen when the variation in $a_{\rm sym}$ is
considered (Fig.~\ref{fig:Tvary}, right). There is a dramatic change of the
isospin-dependent terms, namely the density-dependent symmetry-energy
coefficient $a'_{\rm sym}$ and the surface-symmetry coefficient $a_{\rm
ssym}$. The magnitude of variations in $a_{\rm surf}$ is
comparable to that from the first set. That is, a lower value of
$a_{\rm sym}$ can compensate for a larger effective mass. (A similar
conclusion, in a slightly different context, has been drawn in
Ref.~\cite{[Agr04]}.)  And here is the first time that we see a handle
on the curvature coefficient $a_{\rm curv}$: it weakly decreases with
$a_{\rm sym}$.

The above  trends explain most of the results for the standard Skyrme
functionals displayed in Table~\ref{tab:coeffinout}. However, the
comparison between BSk1 and BSk6 leaves some puzzles. The functional
BSk6 has a lower effective mass than BSk1 (0.8 versus 1.05), and yet, the 
surface-energy coefficient slightly  shrinks. This can probably be related to
an improved treatment of the center-of-mass correction in BSk6.

As already mentioned, there exist different prescriptions for defining
the spin-orbit interaction in SHF \cite{[Ben03]}. One variation concerns
the possible contribution from the kinetic forces resulting in
a $\tensor{J}^2$-term in the energy functional ($\tensor{J}$ 
is the spin-orbit tensor). We have studied this point within our systematic
calculations. The results in Fig.~\ref{fig:Tvary} employ a version
without (left panel) and with (right) that term. The same conditions
are met for the left set at \mbox{$m^*/m=0.81$} and the right set at 
\mbox{$a_{\text{sym}}=32$} MeV. All quantities shown are insensitive 
to that change, except for the density-dependent symmetry energy coefficient
$a'_{\rm sym}$, which exhibits a surprisingly large sensitivity to 
$\tensor{J}^2$. We have checked this point in more detail
and concluded  that the spin-orbit  tensor term  basically changes the offset
of $a'_{\rm sym}$ but has no influence on its trends. Anyway, it is
noteworthy that a change of shell structure (here, via the spin-orbit
interaction) can have such a dramatic effect on a bulk property
of the functional.

%
%
\section{Conclusions}
\label{conclusions}

In this study, we  present a systematic survey of nuclear
surface properties in terms of the liquid drop model. Surface,
surface-symmetry, and curvature energy coefficients  are deduced as
they are defined in the LDM, namely from the trends $\propto A^{-1/3}$
and $\propto A^{-2/3}$ in the binding energies of finite nuclei over a
wide range of sizes. In order to achieve sufficiently precise values,
we have evaluated a smooth background of binding energies by
subtracting the shell corrections, and we considered  huge nuclei
containing  up to $10^5$ nucleons.

Our calculations  show how the bulk-matter limit is recovered in
finite nuclei. While it has been known from earlier studies
\cite{[Boh76a]} that semi-classical features are revealed only for
nuclei with \mbox{$A > 5000$}, we found that extremely massive  nuclei
are essential  in order to pin down unambiguously the macroscopic
surface-related parameters. The question emerges what role the LDM
background plays  for actual  nuclei which are extremely small at that
scale. 

What is the influence of uncontrolled residual shell effects on LDM
parameters when only dealing with small sample  of actual nuclei? The
recent SHF work \cite{[Sat05]} that used   a sample of ``small" nuclei
to extract  the symmetry-energy and surface-symmetry energy
coefficients can provide a hint. While in some cases their results for
$a_{\rm ssym}$ are close to ours (e.g., they obtain $a_{\rm
ssym}$=--49.2 MeV for SkM$^*$ versus --52 MeV here), there are forces
for which the difference is fairly large (e.g., they obtain $a_{\rm
ssym}$=--49 MeV for SkO while we get --58 MeV), and no clear tendency
can be observed.

In fact, the importance of shell effects depends on the observable.
Energies as such exhibit a quick quenching of shell effects with
increasing $A$ (see Fig.~\ref{fig:testhell}). However, energy
differences are required when evaluating the effective surface energy.
As illustrated in Fig.~\ref{fig:asurf}, these differences reveal shell
effects up to any size. This  annoying feature limits the
precision with which one can deduce the higher-order  LDM parameters,
in particular  the isovector parameters which are 
roughly  determined empirically. Those  are also much harder to pin
down in the present analysis, and probably in any analysis, as the
available span of $I$ in bound nuclear systems is rather small.

The leading LDM parameters can be determined sufficiently well as to
allow a thorough analysis.  First, we have studied a broad selection
of widely used ``standard'' SHF and RMF parameterizations. We find
that all isoscalar energy coefficients, including surface and
curvature ones, show only a few percent variation, while the isovector
energy coefficients might differ by a factor of two or even more between
forces. In the second step, we have worked out some interrelations by
a systematic variation of forces. We reconfirm that isovector features
are much more sensitive  to parameter changes than the isoscalar
terms. We have also spotted a surprisingly strong interrelation
between the spin-orbit force and some LDM properties (the density
dependent asymmetry energy coefficient). This demonstrates  that the
shell structure and smooth LDM background are intimately connected.
The variation of the LDM   coefficients does not, however, directly
translate into similar variations of the corresponding energies.

Finally, we note  that in this work we follow  a strictly
``empirical'' approach  relating (shell-corrected) binding energies to
the LDM parameters.  One can amplify (and extract)  some of the LDM
parameters better by  using other, more sensitive, observables, such
as fission barriers or  energies of superdeformed states. Such studies
are currently being pursued.

%
%
\begin{acknowledgments} 
This work was supported in part by the National Nuclear Security
Administration under the Stewardship Science Academic Alliances
program through the U.S. Department of Energy Research Grant
DE-FG03-03NA00083; by the U.S. Department of Energy
under Contracts Nos. DE-FG02-96ER40963 (University of Tennessee),
DE-AC05-00OR22725 with UT-Battelle, LLC (Oak Ridge National
Laboratory), DE-FG05-87ER40361 (Joint Institute for Heavy Ion
Research), DE-FG02-00ER41132 (Institute for Nuclear Theory), 
W-31-109-ENG-38 (Argonne National Laboratory);
by the National Science Foundation under Grant
No. PHY-0456903;
 by the Bundesministerium f\"ur Bildung und Forschung (BMBF),
Project Nos.\ 06 ER 808 and 06 ER 124; 
and and by the Hungarian OTKA fund nos T37991 and T46791.
\end{acknowledgments}
%


\end{document}